# Inhomogeneous low temperature epitaxial breakdown during Si overgrowth of GeSi quantum dots

C.W. Petz and J.A. Floro
Department of Materials Science and Engineering
University of Virginia, Charlottesville, VA 22904

**Abstract:**

Low temperature epitaxial breakdown of inhomogeneously strained Si capping layers is investigated. By growing Si films on coherently strained GeSi quantum dot surfaces, we differentiate effects of surface roughness, strain, and growth orientation on the mechanism of epitaxial breakdown. Using atomic force microscopy and high resolution cross-sectional transmission electron microscopy we find that while local lattice strain up to 2% has a negligible effect, growth on higher-index facets such as {113} significantly reduces the local breakdown thickness. Nanoscale growth mound formation is observed above all facet orientations. Since diffusion lengths depend directly on the surface orientation, we relate the variation in epitaxial thickness to low temperature stability of specific growth facets and on the average size of kinetically limited growth mounds.

**Introduction:**

Investigations into the low temperature epitaxial growth of Group IV semiconductors have provided useful insights on surface-mediated mechanisms for breakdown of the crystalline structure under conditions of limited adatom mobility. Existing research in this area has examined Si homoepitaxy[1,2], Ge homoepitaxy[3-6] and to a lesser extent, strained Ge/Si heteroepitaxy[7]. In this paper, we examine low-temperature growth of Si "capping layers" on GeSi/Si (001) self-assembled quantum dots. Although this effort is primarily driven by fundamental considerations, it is also partly motivated by our associated research on directed self-assembly of ultra-small Ge quantum dots (QDs) on prepatterned Si substrates, where interdot spacings as small as 22 nm have been achieved[8]. In order to suppress coarsening processes that lead to inhomogeneity in the dot distribution, it is necessary to use low temperature growth and capping. This raises the question of how epitaxial breakdown processes are modified for low-temperature growth of Si over quantum dots, which notably present different growth facets, inhomogeneous misfit strain, and a pre-roughened surface morphology on the mesoscale. While low temperature homoepitaxial growth has been extensively studied by others on a variety of semiconductor surfaces, an investigation of the epitaxial breakdown interface on inhomogeneous surfaces should contribute to a more fully-developed understanding of the mechanisms for breakdown.

A variety of mechanisms for epitaxial breakdown have been suggested, including the role of defect accumulation[9], continuous breakdown[10], hydrogen absorption[11,12], and kinetic roughening[1,3-6,13-16]. A common picture emerging from these studies is that during low temperature homoepitaxial growth on (001) surfaces of Group IV semiconductors, {111} facets are eventually exposed at the growth surface upon which extensive faulting can occur, directly leading to breakdown of epitaxy. Here we utilize well known GeSi/Si(001) QD nanostructures as a 3D canvas for low temperature epitaxial overgrowth of Si at 160°C. This system allows us to examine the roles of both inhomogeneous strain and local island faceting on the epitaxial breakdown process.  A similar study was recently reported where Si overgrowth was carried out at higher temperatures than used here, resulting in larger breakdown thicknesses[17].  That work attributed breakdown to fault generation on {111} through strain-induced partial dislocation introduction.   By reducing the Si growth temperature, we can better pinpoint the localized regions for the initiation of breakdown during overgrowth.  We find that breakdown occurs much earlier over {113} island facets, but other facets such as {105} do not affect the breakdown relative to {001}, and perhaps even augment the critical thickness.  In addition, no effect of local strain variations on the epitaxial breakdown thickness are observed here.

Bratland, et al., have recently ascribed kinetic roughening effects as the primary mechanism for {111} faceting and eventual epitaxial breakdown during Ge homoepitaxial growth[6]. They showed that shallow growth mounds form due to the presence of Erlich-Schwoebel (E-S) barriers and where mound intersection occurs, local {111}-faceted cusps form.  Extensive faulting occurs during subsequent growth on these {111} cusps, followed by an abrupt transition to the amorphous structure at larger thicknesses.  They define $h_1$ as the mean initial thickness at which defect generation begins, corresponding to the formation of {111} cusps, and $h_2$ as the mean thickness at which the layer has become fully amorphized. We also observed nanoscale mound formation on the surface of our Si cap layers, at length scales much smaller than the buried quantum dots, that appear to be intimately linked to epitaxial breakdown.

**Experiment:**

$Ge_{0.5}Si_{0.5}$/Si(001) QDs were grown via ultra-high vacuum molecular beam epitaxy (MBE) (base pressure = $10^{-10}$ Torr). Although we have examined low temperature growth on pure Ge quantum dots, alloy dots were grown for this study to provide larger islands that facilitate detailed transmission electron microscope observations of the breakdown interface. Prior to insertion to the MBE, Si wafers with a miscut of 0.1∘, were chemically cleaned via a standard IMEC/Shiraki process to remove hydrocarbon and transition metal impurities, creating in the final step a passive $SiO_x$ layer. The substrates were outgassed in the MBE at 600°C for > 4 hrs, ramped to 850°C over 30 min to desorb the oxide layer, and cooled to 740°C for deposition of a 50 nm Si buffer layer. Throughout this process, the surface structure was monitored with reflection high energy electron diffraction (RHEED) to ensure 2×1 surface reconstruction and a smooth surface as indicated by the presence of a Laue ring of diffraction spots. We deposited Ge and Si via magnetron sputtering in 3 mTorr of getter-purified Ar. Once a clean surface was obtained, GeSi heteroepitaxy proceeded at 740°C via co-deposition of Ge and Si with a total flux of 0.3 Å/s to a thickness of 31 Å.  The surface was then cooled to 160°C for 30min in UHV for Si capping at 0.1 Å/s.  The capping temperature of 160°C was estimated based on prior thermocouple-based calibrations of temperature vs. heater current.  This growth temperature was chosen to provide a measurable epitaxial breakdown thickness ($h_1$) relative to Si(001) homoepitaxy which was experimentally determined by Eaglesham to be approximately 30 nm[1].  We examined two Si cap thicknesses, 12 and 85 nm.

Ex-situ atomic force microscopy (AFM) was performed with an NT-MDT Solver Pro-M using NSG10 tips with radius <10nm.   Transmission electron microscopy (TEM) and electron energy loss spectroscopy (EELS) were performed on an FEI Titan 80-300 operated at 300kV.

Cross sectional TEM foils were prepared by mechanical thinning to a thickness of <50um and ion milling with 4keV Ar+ at a 5° incident angle to perforation. Samples were examined in the <110> zone axis.

**Results:**

AFM topography scans with representative line profiles are shown in Figure 1, comparing morphology of samples with and without low temperature Si caps. For our growth conditions, island areal density is ~60 um$^{-2}$, showing a clear bimodal distribution of Stranski-Krastanow "pyramids" and "domes" with average diameters of 140 nm and 160 nm, respectively. The pyramid and dome morphologies, which form in order to reduce the compressive biaxial lattice mismatch strain, are ubiquitous in $Ge_xSi_{1-x}$ alloy QD growth[18,19]. Pyramids are bound by {105} facets, while the domes are bound by {113}, {15 3 23}, and {105} facets[20]. AFM shows that a 12 nm thick Si cap grown at 160°C leads to some broadening of the surface features, but most of the representative surface angles are retained. Furthermore, the roughness of the wetting layer regions between the dots, and on the dot facets appears to be identical for the capped and uncapped samples. Hence the surface of the Si cap layer is almost completely conformal to the underlying quantum dot array, despite the occurrence of partial amorphous breakdown as we will demonstrate below. We have also found this to be true for complete epitaxial breakdown of low temperature Si and Ge overlayers on Ge/Si(001) and GeSi/Si(001) quantum dot surfaces.

To examine how epitaxial breakdown and amorphization of the Si cap correlate with the underlying quantum dots, we employed a defect-sensitive organic peracid etching (OPE) technique[21,22]. OPE consists of premixed $CH_3COOH:H_2O_2$ (3:1) and HF. This solution creates peracetic acid that behaves as a weak oxidizing agent. The oxidation rate is determined by the $H_2O_2$ concentration and subsequently limits the rate of material removal by HF. The mechanism of accelerated material removal at defect sites is based upon increased potential energy due to missing bonds, impurities, and dislocation strain fields. Selectivity for defective Si is only about 2x over that of perfect Si, which has an approximate etch rate of 3 nm/min, so some etching of "good" material is unavoidable. Figure 2 shows the OPE etched surface of the Si capped sample after 1 minute in solution. The Si cap over the pyramids and wetting layer exhibits modest, relatively uniform etching. But the Si over the domes shows significant, inhomogeneous etching, indicating the localized formation of defective and/or amorphous structure in these regions. In particular, the linescan comparison shown in Fig. 2 demonstrates that etching was fastest over regions roughly over the {113} facets of the buried dome clusters. The etched GeSi dome profiles exhibit sidewall angles of 40° although measurement of such steep angles is limited by the finite radius of the AFM probe.

To provide a detailed microscopic view of the defective epitaxial breakdown interface, we performed cross-sectional TEM and EELS analysis. For the 12 nm thick low temperature Si capping layer, for which RHEED indicates partial amorphization, we observe defect free epitaxial growth above the (001) wetting layer and the {105}-faceted pyramids[23], as shown in Fig. 3. In agreement with AFM, the Si cap appears perfectly conformal to the underlying pyramid. Epitaxial breakdown is observed to occur above dome clusters, as shown in Fig. 4, but the breakdown front is localized. Similar breakdown morphologies were observed over all 8 domes surveyed in the XTEM specimen. Underfocused bright-field imaging in Fig. 4(a) shows a buried dome, where the position of the Si surface is readily identified (black arrows) by examining through-focus conditions. The XTEM indicates that the Si cap surface is again conformal to the buried dome, in agreement with AFM.  Si EELS mapping (not shown) confirms this result.  Fig. 4(a) shows that epitaxial breakdown occurs along the side facets of the dome island, in agreement with the OPE results of Fig. 2.  Note that at the apex of the GeSi dome, the crystallinity of the 12 nm cap is fully retained.

A high-resolution TEM image of the breakdown interface is shown in Fig. 4(b) and shown outlined in Fig. 4(c).  The breakdown interface is composed of alternating {111} and {001} facets, with an average slope to the interface of about 25°, corresponding to an overall {113} facet.  This correspondence suggests that Si overgrowth on the prominent {113} facet of the GeSi dome cluster is where epitaxial breakdown first nucleates. Reduced epitaxial breakdown thicknesses on {113} surfaces have been observed by others[24,25].

To examine the complete transformation to defective epitaxy ($h_1$) across the sample, an 85 nm Si cap was grown at the nominally identical temperature of 160°C.  Figure 5 shows a cross-section micrograph of typical pyramid and dome islands.  Breakdown begins over the wetting layer at a thickness $h_1$ = 47 nm above the planar wetting layer regions, and above the {105} faceted pyramids. This breakdown thickness is also retained at the dome perimeters.  However, Fig. 5 clearly shows that breakdown occurs earlier over the dome {113} facet (as shown for the 12 nm cap), while over the apex of the dome, $h_1$ is estimated to be 55 nm, even larger than over the wetting layer regions.  It must be acknowledged that the latter estimate is challenging due to the complex contrast in this region of the XTEM specimen.

Our AFM measurements of the 85 nm thick Si cap surface (not shown) indicate that the cap conformally replicates the underlying islands, with no increase in local-scale roughness.  However, Fig. 5 demonstrates that the surface of the thick cap actually exhibits a fine-scale scalloped morphology indicating growth mounds have developed at this thickness.  These mounds have a lateral size of about 12 nm over the wetting layer regions, with a peak-to-valley height of 2-3 nm.  Such fine-scale but high-aspect features were not detected by the AFM tip, which had a nominal 10 nm radius. Further, although more difficult to visualize, it does appear that mounds are forming on the Si cap over the dome clusters as well, but these mounds appear to be even smaller, of order 4 nm lateral size.  The presence of mounds over both the (001) and {113} are also indicated by the cooperative formation of void trails that are readily visible in Fig. 5.

Finally, we observe two abrupt increases in angle of the $h_1$ interface above the dome: from 25° to 54° near the {113}/{105} intersection, and from 54° to 70° directly above the dome apex.  These appear to correlate back to changes in local faceting of the GeSi quantum dots.  The increase to 54° is correlated with a transition from growth over {113} facets to growth over the domes' {105} facets.  The increase in angle towards 70° is then associated with oriented (001) epitaxy which retards impingement of the bounding defective sublayer.

**Discussion:**

Our clear observation of enhanced epitaxial breakdown of Si over the {113} facets of the overgrown GeSi domes is consistent with previous reports on Si homoeptiaxy[25].  In the context of a picture wherein formation of {111} facets, and associated fault generation, is required to nucleate the amorphous phase, the {113} structure could be conducive to exposure of {111} planes.  The unreconstructed {113} surface consists of single atomic terraces of alternating {111} and {001} [26].  This is shown in Fig. 6.  While the Si {113} is a true facet, and is known to exhibit a stable 3x2 reconstruction at lower temperatures[27], we will assume for simplicity that the reconstruction is broken during low temperature growth. In Fig. 6, a single monolayer-height step is shown, which generates a 2-unit wide {111} facet. Hence, any local roughening of the {113}, e.g., step bunching or mound formation, naturally generates extended {111} facets. While this simple picture provides an appealing qualitative explanation for why epitaxial breakdown thickness is reduced on the {113}, we cannot say for sure that the reconstruction has been broken during low temperature growth of Si over the GeSi island. It is noted, however, that any tensile strain in the Si cap layer growing over the partially-relaxed island should contribute towards destabilization of the 3x2 reconstruction due to its large inherent tensile bond strain.

Bratland, et al., linked nucleation of the amorphous phase to the formation of growth mounds having a critical aspect ratio. The critical ratio was related to a peak-to-valley distance that is larger than the temperature-dependent diffusion length.[6,14] They attributed mound formation to the presence of E-S barriers on the crystalline Ge surface, although their mounds tended to be much larger, and occurred in much thicker films, than observed in our case. We also observe breakdown coupled to mound formation on the Si (001) surface that appears qualitatively quite similar to their work. The presence of E-S barriers on Si (001) is not established, although step bunching and mound formation have been observed previously and attributed to alternative roughening mechanisms[13,28,29]. Additionally, we note that growth mound formation has been observed on *fully amorphous* Ge, Si and metal alloy films[30,31]. In this work we also observe mound formation on {113}. The smaller size of the mounds implies reduced overall diffusivity on this surface and correlates with the smaller breakdown thickness.

We find that the growth mounds are accompanied by void trails (see Fig. 5), as was observed previously[6]. Void trail formation is intimately linked to local surface roughening and mound formation[30,32]. The trails are tilted by about 15° with respect to <001>. Similarly, the surface-replicas of the huts and domes are all offset in the same direction relative to the underlying GeSi islands, and in the same direction as the void trails, but at an angle of 23°. In the limit of zero adatom mobility, the tilt angle of the void trails relative to the film plane should equal that of the incident flux (in our case, 30°). That the tilting of the mounds is considerably smaller than the incidence angle of the Si flux suggests that some surface transport over the nanoscale mounds is occurring. The larger tilt angle of the GeSi island surface-replicas is consistent with relatively reduced transport due to the much larger length scale of these features.

Strain does not appear to affect the epitaxial breakdown process in these films. We note that the Ge wetting layer and both the pyramid and dome islands are fully coherent. Growth of Si over the wetting layer will not impose any elastic strain in the Si cap. However, over the islands, which partially relax due to their 3D geometry, there will be local strains imposed on the Si. Continuum elastic modeling indicates that the apex of the dome is expected to exhibit almost complete strain relaxation [33-35], and therefore the Si cap should be strained up to 2% tensile when it overgrows the apex. Another region of potentially large strain in the overgrown Si cap would be above the perimeter of the dome cluster, where the dome and the Si substrate are under excess compression. Similar, but smaller, strains will be imposed in the cap by the pyramids. Careful inspection of several domes and pyramids in our specimen indicates that Si breakdown is not occurring over the apices or the perimeters of the underlying islands.

In recent work by Lin, et al., Si was grown over Ge/Si(001) QDs at 300°C, resulting in $h_1$ ≈ 30 nm over the QDs and, we estimate, 100 nm over the wetting layer[17]. Such thicknesses are much larger than observed here due to their higher growth temperatures. They observed stacking faults localized over the buried QDs that appeared to originate at the perimeters of the Ge dome islands, where compressive stress is present in the Si cap. They attributed the formation of faults to passage of partial dislocations due to the stress. We did not observe this breakdown mode, perhaps due to the lower growth temperature used here, where growth mounding and roughening, especially on the {113}, promotes breakdown before the critical thickness for shear-related mechanisms. Also, the Ge content in our islands, and hence the strain in the Si cap is smaller in our experiments than that for Lin, et al. However, though we observe no direct correlation of strain on $h_1$, we note that strain does impact the relative stability of surface reconstructions and adatom diffusivity; thus a hybrid picture of epitaxial breakdown involving kinetic- and strain- effects is required.

In conclusion, during low temperature Si overgrowth of GeSi coherently strained islands and wetting layer, we show that the low temperature epitaxial thickness, $h_1$, depends primarily on the mesoscopic facet orientation of the Si, which is conformally inherited from the islands. Globally, epitaxial breakdown occurs earliest over the {113} facets due to the ease of creating local {111} surfaces associated with step formation. We observe kinetically limited growth mounds on all QD related facets and note that the mean mound width is directly related to the epitaxial thickness and thus to the local surface diffusivity. The small mound size on {113} vis-à-vis {001} implies reduced diffusivity on this surface, further enhancing the tendency to breakdown. Finally, we show that $h_1$ for Si is independent of coherent strain of at least 2%, suggesting that at these low strain levels the initiation of defects is dominated by kinetic growth mound formation.

**Acknowledgments:**

The authors would like to thank Prof. James Howe and P. Palanisamy for their guidance with electron microscopy and J.C. Duda for his assistance with image analysis. Support from the United States Department of Energy Office of Basic Energy Sciences is gratefully acknowledged under grant number: DE-FG02-07ER46421.

**List of figure captions:**

Figure 1: AFM topography images and associated linescans of typical pyramids (black dashed line) and domes (solid blue line). The red dashed lines on the graphs are associated with the local surface angles of the indicated domes in the [110 azimuth]. (a) uncapped $Ge_{0.5}Si_{0.5}$ quantum dots. (b) Morphology after low temperature growth of a 12 nm Si cap, where partial amorphization has occurred as indicated by RHEED and TEM.

Figure 2: AFM topography and associated linescans of SiGe islands with a defective epitaxial Si cap after OPE etching for 1 minute. A typical hut (black dashed line) and dome (solid blue line) are shown along with the local surface angles of the etched dome.

Figure 3: TEM bright field cross-sectional image of a buried GeSi pyramid and its associated Si EELS map, for the sample capped with 12 nm Si at 160°C. The mottled contrast in the bright field image is due to specimen thinning artifacts.

Figure 4: Cross-sectional TEM of a GeSi dome, arrows in (a) indicate the amorphous Si-cap free surface in underfocused conditions (Δf=-500nm). HRTEM of the crystalline-amorphous interface from the indicated box is shown in (b) and a corresponding sketch of this interface is shown in (c).

Figure 5: TEM bright field cross-sectional image of a buried GeSi pyramid and dome. The sharp contrast features on the far right are associated with a bend contour of the thin specimen and are not related to the defective epitaxial region.

Figure 6: Crystallographic orientation of a Si {113} surface. The dashed line follows the average {113} terrace surface, while the heavy black line delineates local {111} and {100} segments. A single monolayer-height step is indicated by an "S" illustrating the ease of generating extended {111} facets.

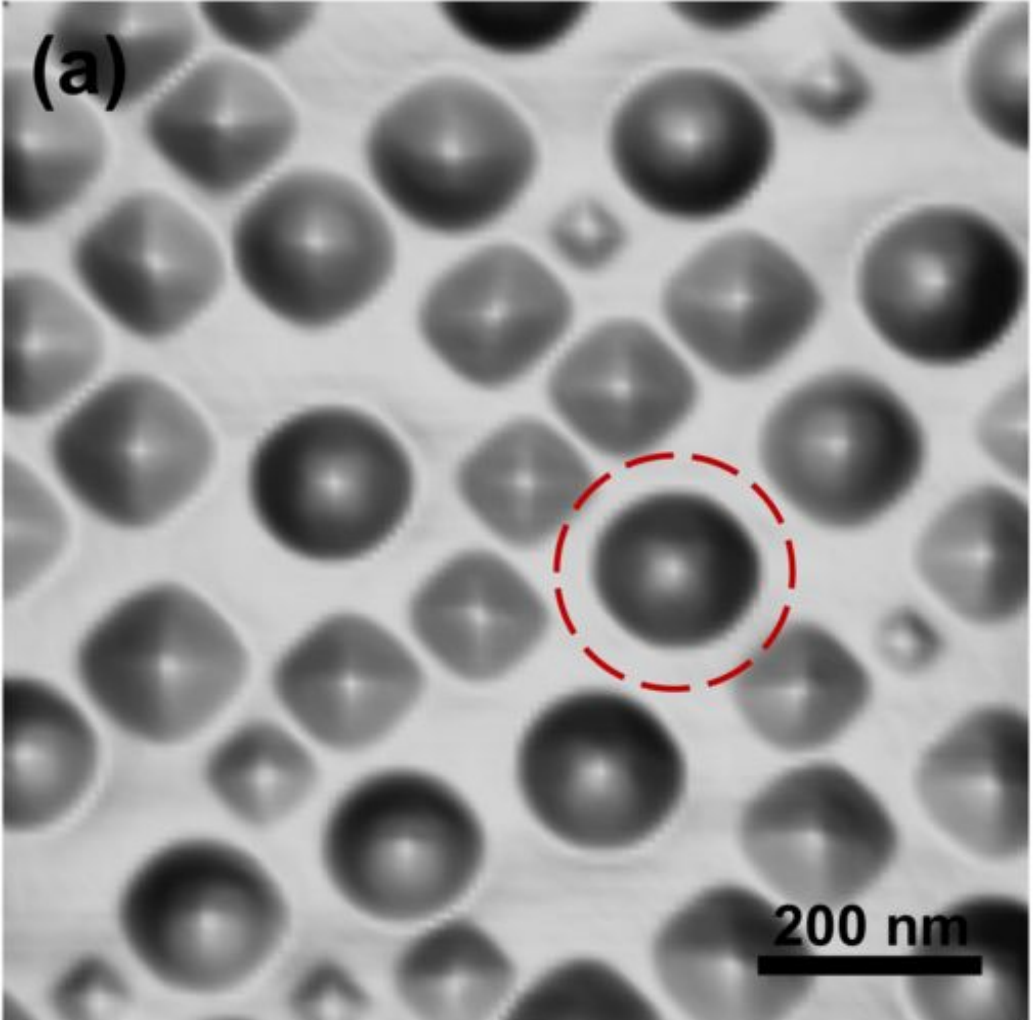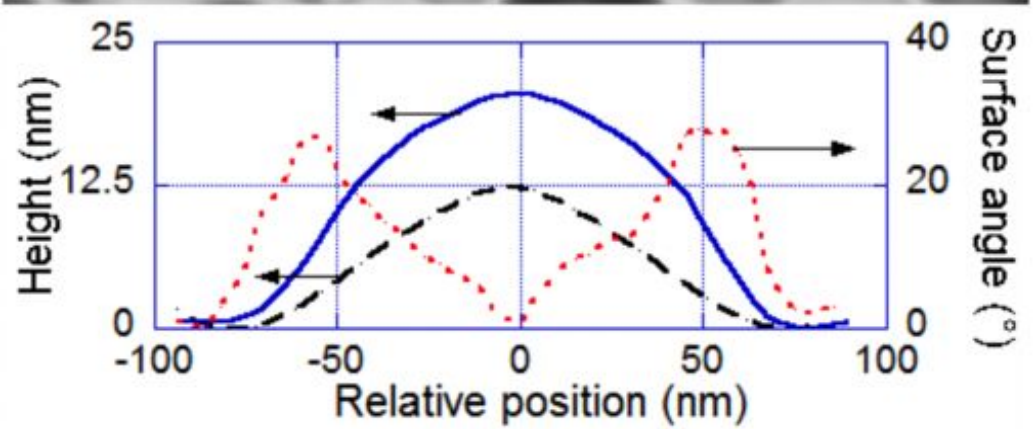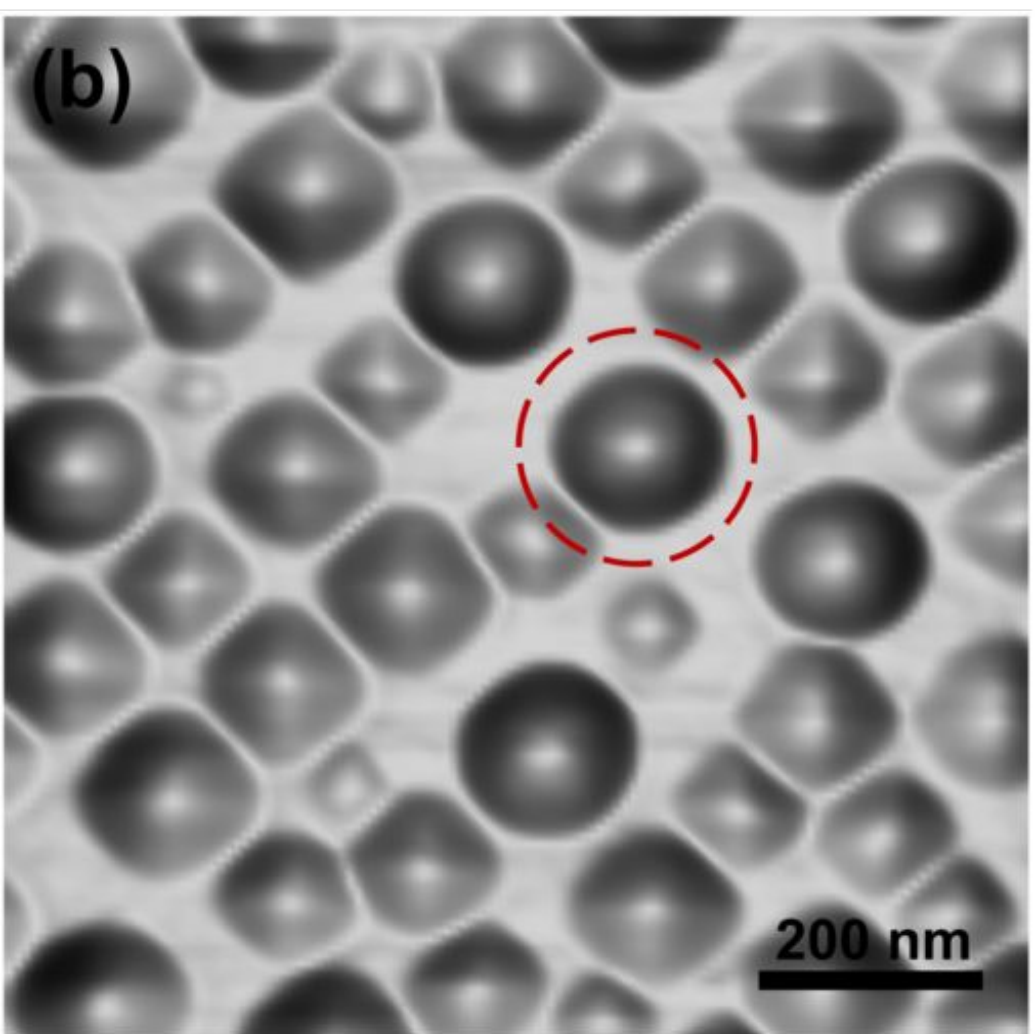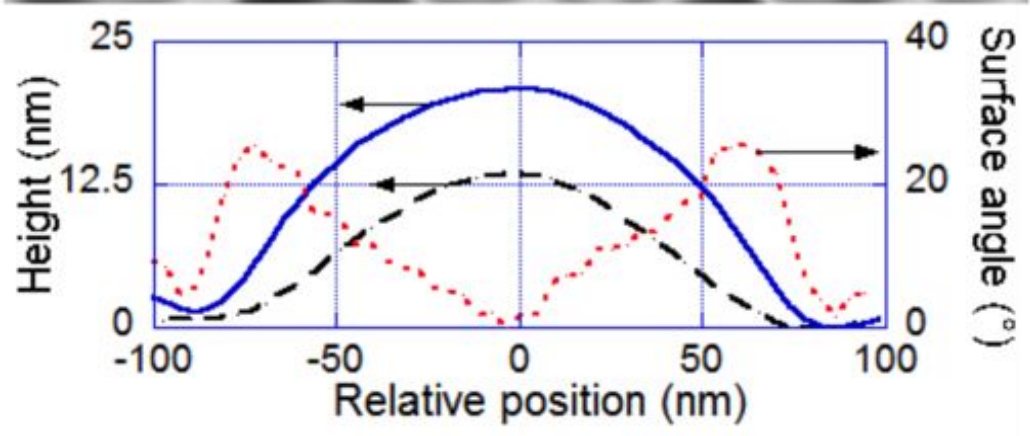

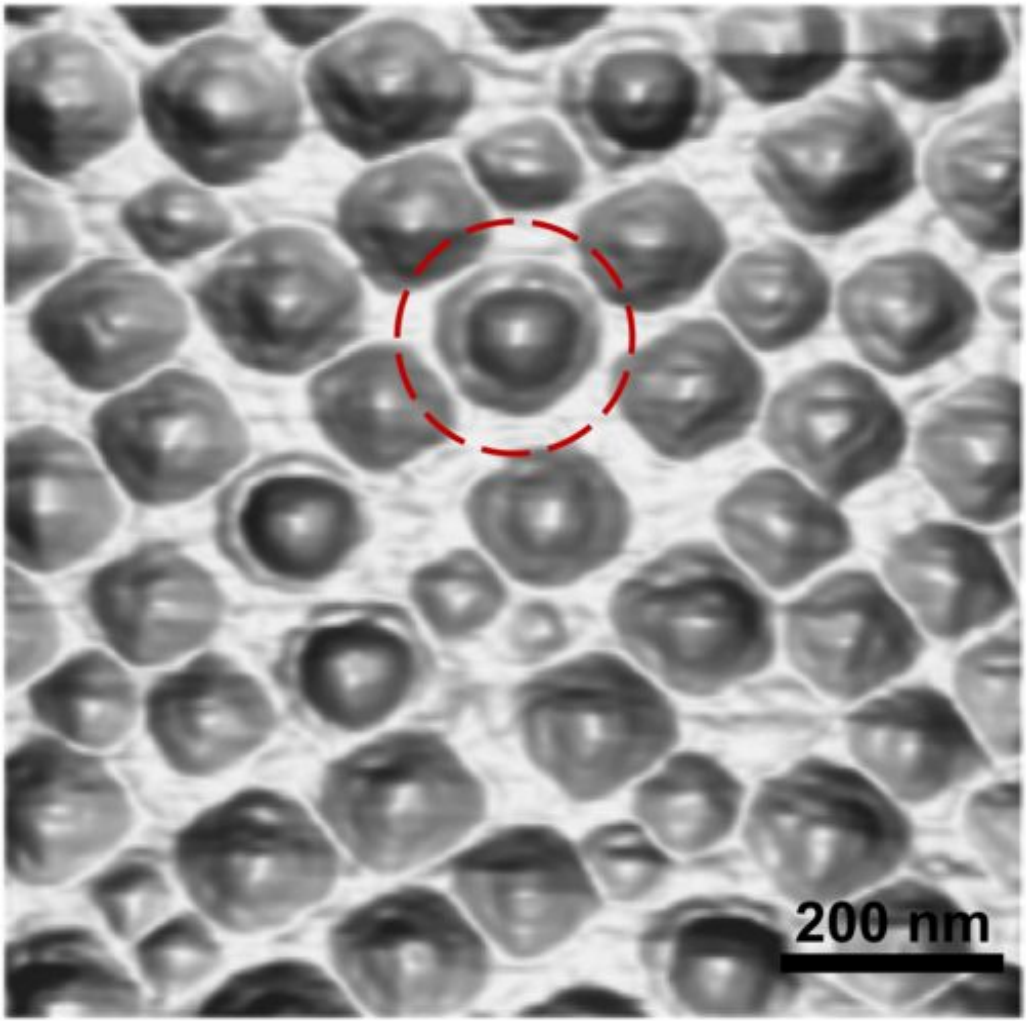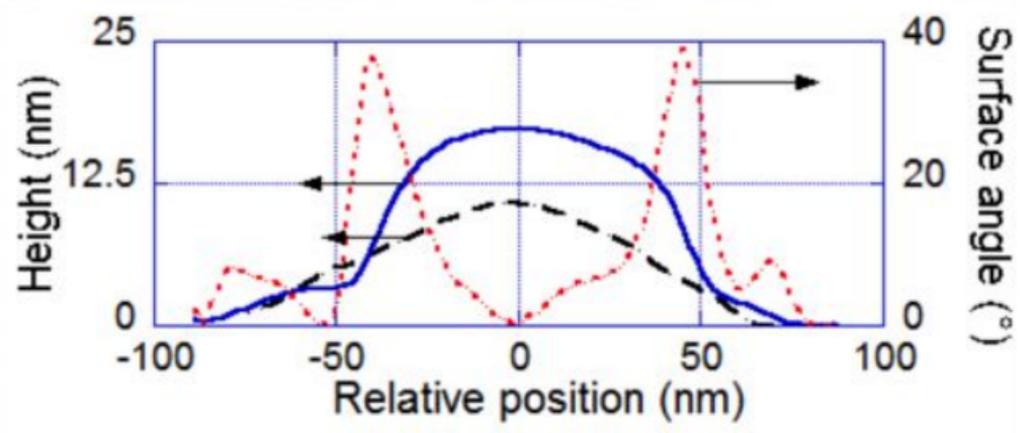

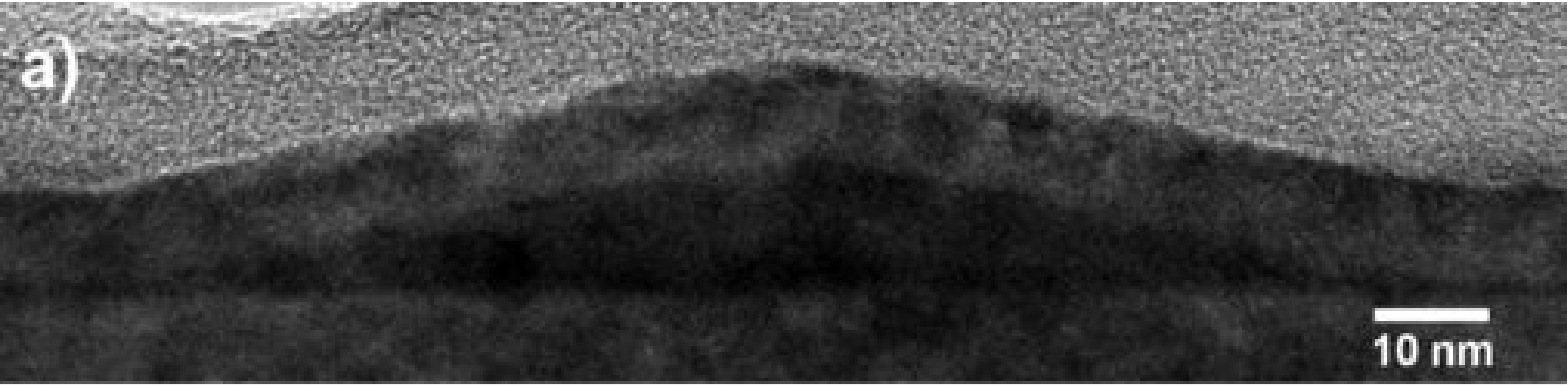
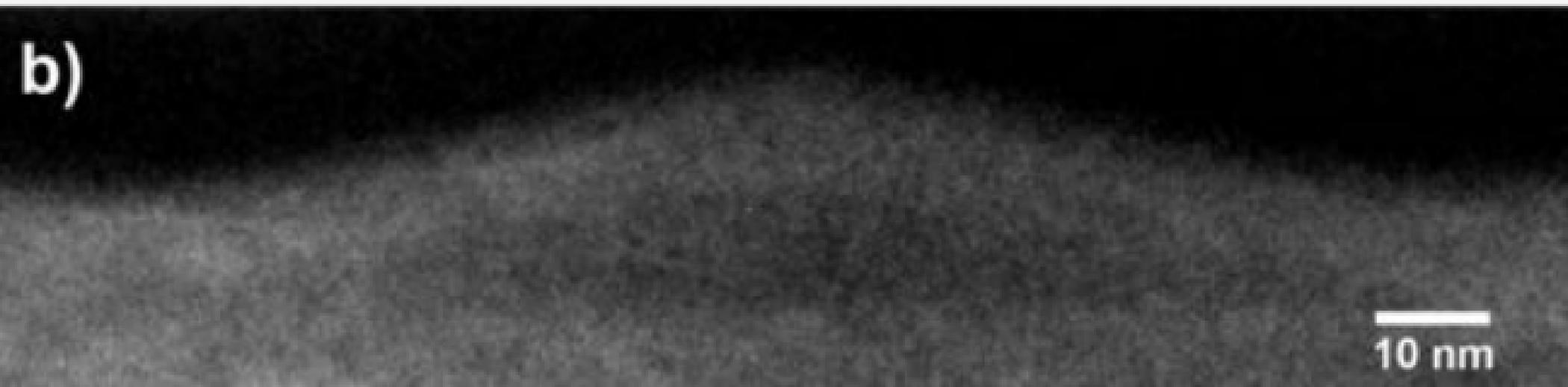

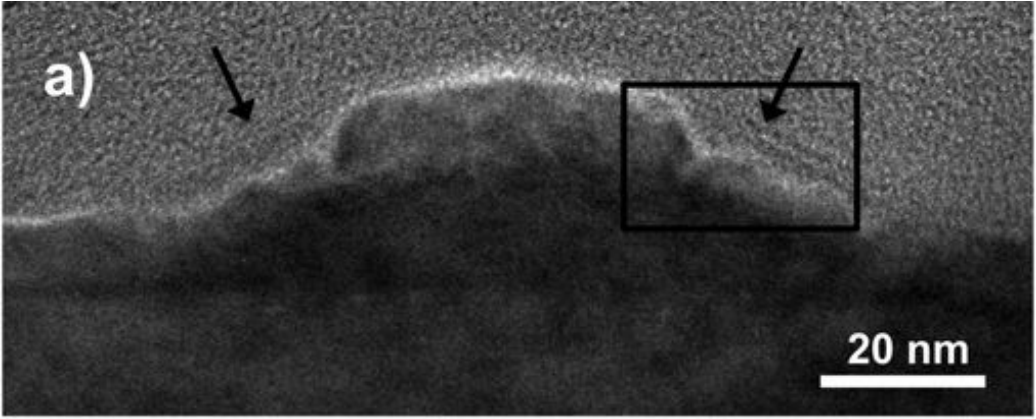
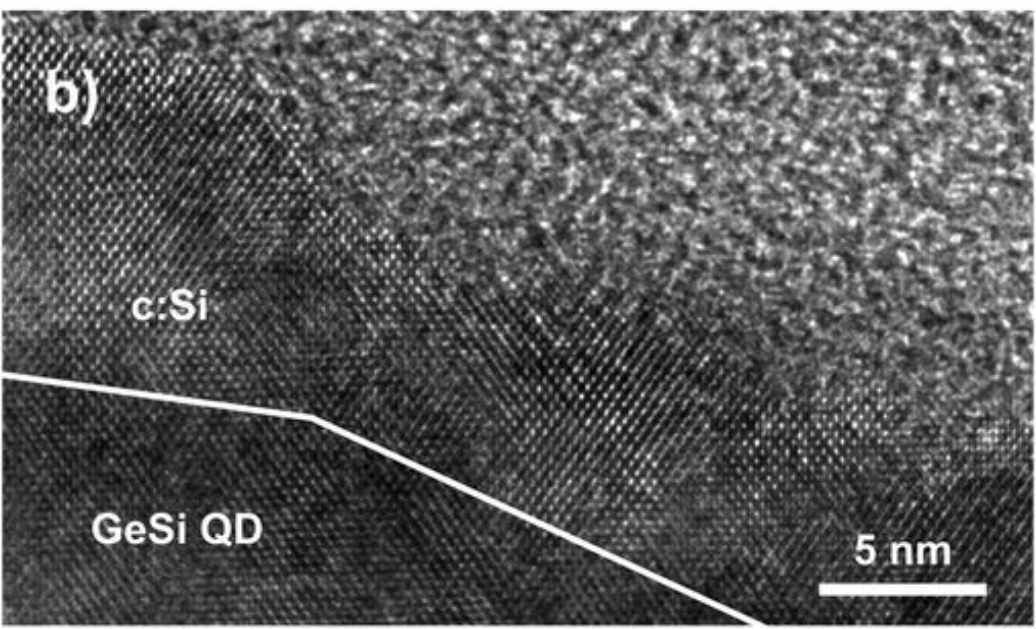
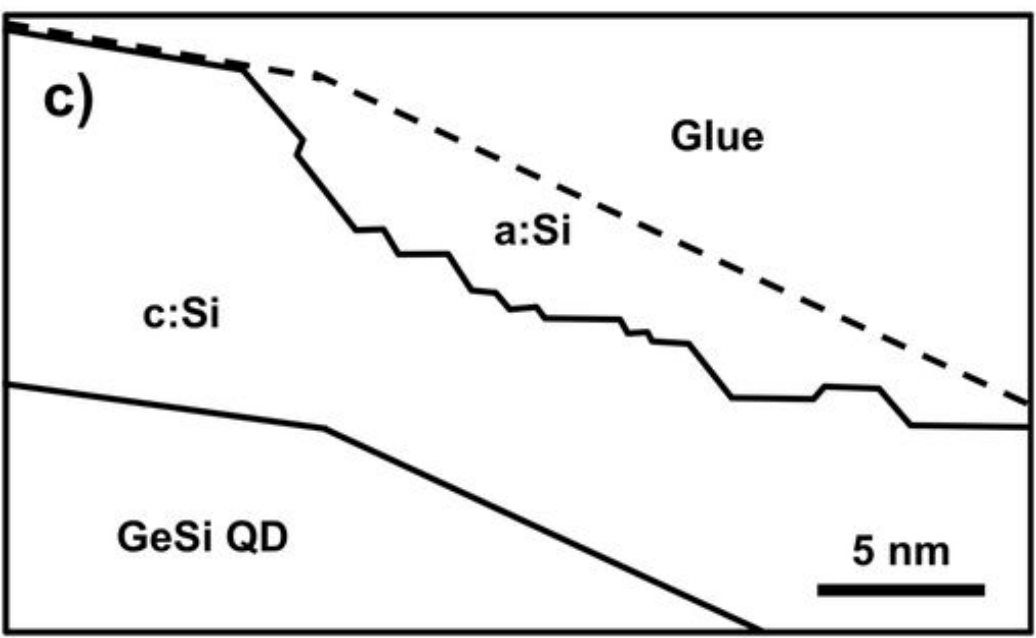

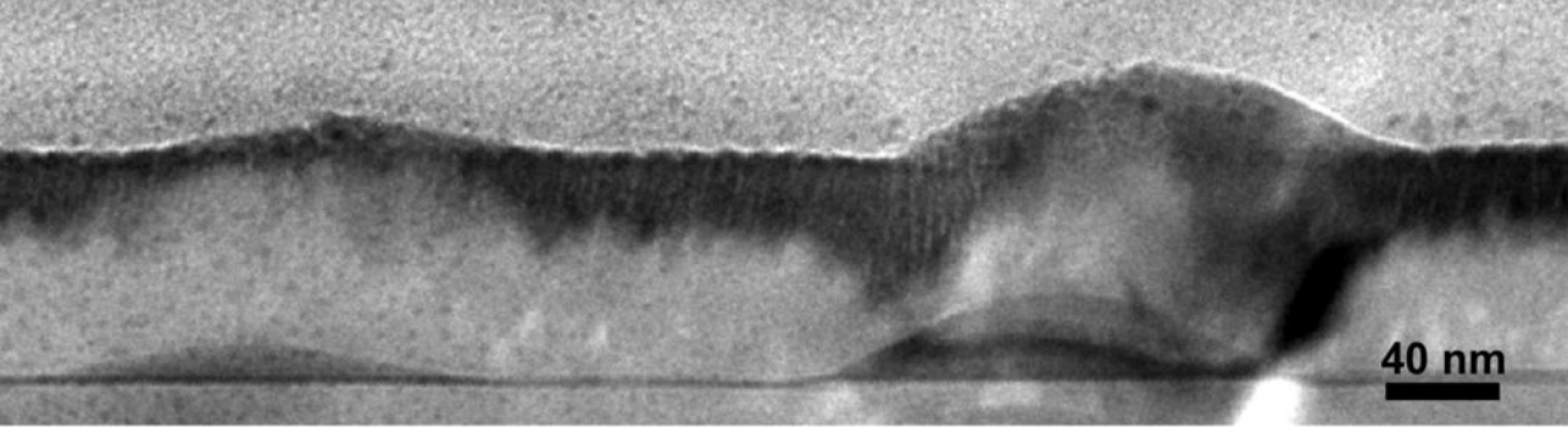

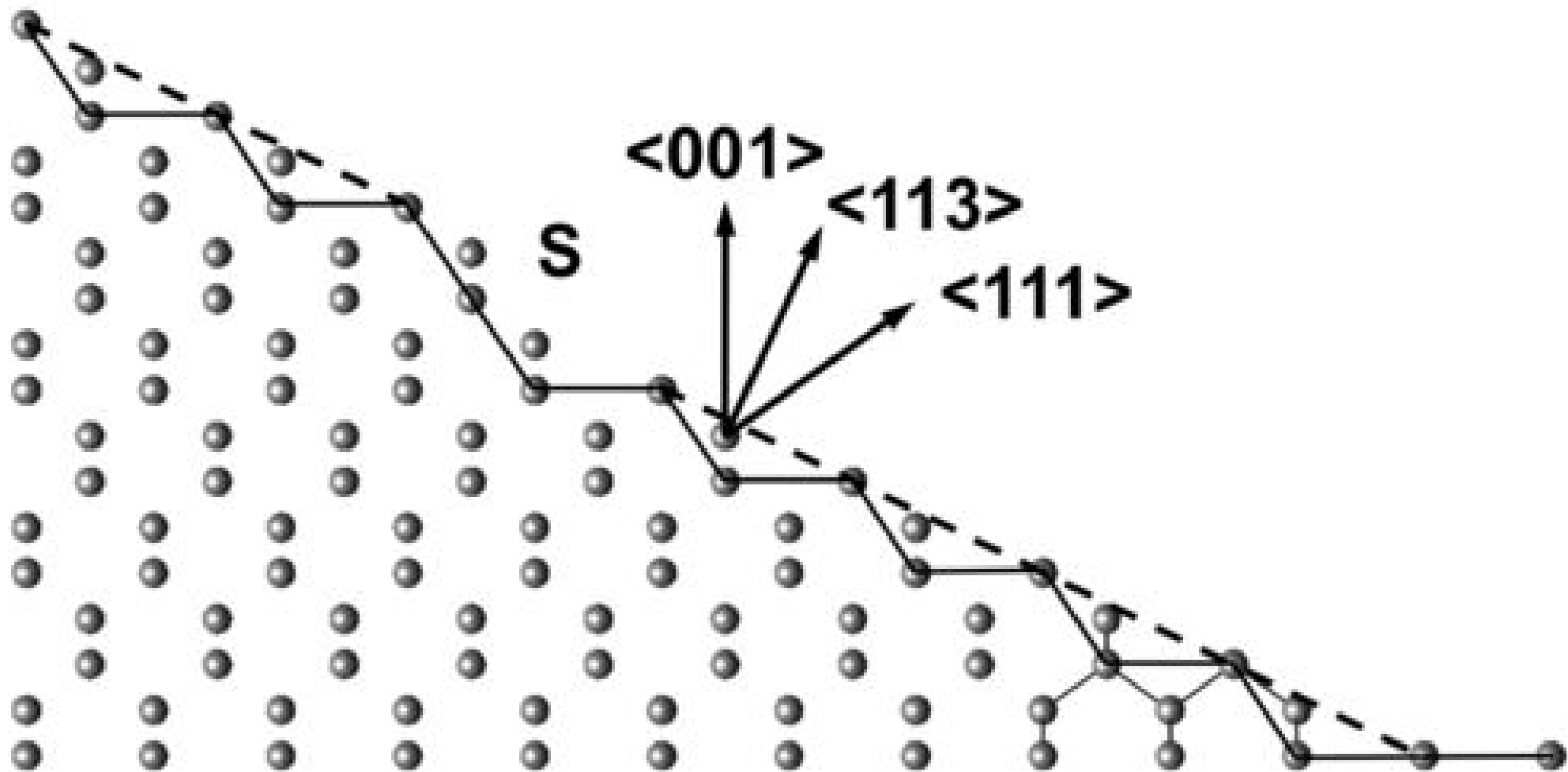